# Gate voltage modulation of spin-Hall-torque-driven magnetic switching

Luqiao Liu[1], Chi-Feng Pai[1], D. C. Ralph[1,2] and R. A. Buhrman[1]

[1] Cornell University, Ithaca, NY 14853

[2] Kavli Institute at Cornell, Ithaca, NY 14853

**Two promising strategies for achieving efficient control of magnetization in future magnetic memory and non-volatile spin logic devices are spin transfer torque from spin polarized currents and voltage-controlled magnetic anisotropy (VCMA). Spin transfer torque is in widespread development as the write mechanism for next-generation magnetic memory, while VCMA offers the potential of even better energy performance due to smaller Ohmic losses. Here we introduce a 3-terminal magnetic tunnel junction (MTJ) device that combines both of these mechanisms to achieve new functionality: gate-voltage-modulated spin torque switching. This gating makes possible both more energy-efficient switching and also improved architectures for memory and logic applications, including a simple approach for making magnetic memories with a maximum-density cross-point geometry that does not require a control transistor for every MTJ.**

Improvements in understanding the fundamental physics of the coupling between ferromagnets, spin currents, and electric fields have led to new schemes for achieving electrical control of magnetic devices with orders-of-magnitude better efficiency compared to magnetic-field-based manipulation. Two of the most promising mechanisms are spin transfer torque, in which a torque is applied to a ferromagnet by direct transfer of spin angular momentum from a spin current[1-3], and voltage-controlled magnetic anisotropy (VCMA), in which an electric field alters a ferromagnetic film's perpendicular anisotropy by changing the electronic structure at a ferromagnet/oxide interface[4-10]. Spin torque can readily drive both deterministic bipolar switching[11] and steady-state oscillations[12]. VCMA has been shown to enable strong tuning of coercive magnetic fields[13] and direct toggle switching by voltage pulses[14]. Very recently, an efficient new strategy for generating spin currents to be used for spin torques has been demonstrated, employing the spin Hall effect[15-17] from an in-plane charge current within a heavy metal thin film such as Pt or Ta to generate a vertically-flowing spin current that applies a torque to an adjacent magnetic film[18-21]. In a previous experiment, we demonstrated a 3-terminal magnetic device (Fig 1a) in which the spin Hall torque from a Ta layer is used to switch the magnetic free layer of a magnetic tunnel junction (MTJ) on top of the Ta, while the MTJ is used to read out the magnetic orientation[22]. Here we show that in this 3-terminal structure a voltage applied across the MTJ can, via the VCMA mechanism, produce a large modulation the critical current for spin-Hall-torque switching, thereby providing local gating of switching. This can be used to reduce the energy required for spin torque switching and, more importantly, it provides fundamentally new strategies for implementing magnetic memory and logic functions. We show, *e.g.*, how gate-modulated switching can enable an improved design for magnetic memory with a maximum-density cross-point architecture.

The devices we study consist of a 6 nm thick, 1 µm wide Ta strip with a $Co_{40}Fe_{40}B_{20}$(1.5)/MgO(1.2)/$Co_{40}Fe_{40}B_{20}$(4) (thicknesses in nanometers) MTJ on top, with the MTJ having an approximately elliptical 100 × 350 $nm^2$ cross section with the long axis perpendicular to the Ta strip. Other device dimensions and fabrication procedures are given in Fig. 1a and the Methods section. Figure 1a also shows the directions of electron flow associated with our conventions for positive currents. The resistance of the Ta strip is 4 kΩ. From measurements of the MTJ resistance as a function of in-plane magnetic field (Fig. 1b), we determine that the magnetic layers have predominantly in-plane anisotropy, the resistance-area product of the MTJ is 35 Ω-$µm^2$, and the tunneling magnetoresistance (TMR) is ~17% at zero bias voltage. As in ref. 22, to achieve switching by the spin Hall torque we apply charge current $I_{Ta}$ through the Ta strip to exert a spin torque on the 1.5-nm-thick free CoFeB layer of the MTJ. The phenomena discussed in this Article appeared consistently in five samples.

The motivation behind our experiment is that the critical current density in the Ta for spin torque switching of an in-plane polarized magnetic free layer by the spin Hall effect should (in the absence of thermal activation) depend on the effective perpendicular demagnetization field $H_{demag}^{eff}$ of the free layer as[3,22]

$$|J_{c0}| \approx \frac{2eM_s t_{free} \alpha}{\hbar \theta_{SH}} \left( H_c + 0.5 H_{demag}^{eff} \right), \tag{1}$$

and $H_{demag}^{eff}$ should be tunable as a function the voltage $V_{MTJ}$ applied across the MTJ because of the VCMA effect:

$$H_{demag}^{eff} = 4\pi M_S - 2K_u(V_{MTJ})/M_S. \tag{2}$$

Here $e$ is the electron charge, $M_S$ is the saturation magnetization of the CoFeB free layer, $t_{free}$ = 1.5 nm is its thickness and $\alpha$ the value of its Gilbert damping, $H_c$ is its within-plane magnetic anisotropy field, and $K_u(V_{MTJ})$ is the voltage-dependent perpendicular anisotropy energy coefficient of the free layer. Based on Equations (1) and (2), it should be possible to control the critical current within the Ta layer required for switching the MTJ by applying a gating voltage to the MTJ. In previous measurements, we determined that $d\left(H_{demag}^{eff}\right)/dV_{MTJ} = 730 \pm 90$ Oe/V for the same multilayer stack used for our samples here, based on the change in the spin-torque-excited magnetic precession frequency of the free layer as a function of MTJ voltage[23]. This corresponds to a change in demagnetization energy per unit electric field $|d(K_u t)/dE|$ $= [M_S t_{free} t_{MgO}/2] d(H_{demag}^{eff})/dV = 70$ μJ/m$^2$ (V/nm)$^{-1}$, consistent with previous measurements in similar devices[6,9,13,14,24,25].

To test for the effect of the MTJ voltage on the spin Hall torque switching, we applied voltage pulses across the MTJ simultaneously with current pulses in the Ta microstrip. Large resistors were connected in series with the MTJ to prevent the current applied to the Ta microstrip from flowing through the MTJ (see Fig. 2a and Section S1 of the Supplemental Information). All measurements were performed at room temperature, and a constant external magnetic field of -25 Oe was applied to cancel the dipole magnetic field from the fixed (thicker) magnetic layer in the MTJ. Switching was monitored by recording the differential resistance across the MTJ after each pulse. Figure 2b shows data obtained with pulse lengths of 10 ms. When $V_{MTJ} = 0$, we observe reversible switching between the parallel (P) and antiparallel (AP) states of the MTJ driven by the spin Hall torque with switching currents $I = -0.58$ mA and 0.44 mA. For $V_{MTJ} = -400$ mV, the room temperature spin Hall switching currents are dramatically reduced, by more than a factor of six, to $I = -0.08$ mA and 0.07 mA. However, the switching currents are less sensitive to positive values of $V_{MTJ}$; for $V_{MTJ} = 400$ mV we observe switching currents of $I = -0.55$ mA and 0.47 mA. The full dependence of the room-temperature switching currents on $V_{MTJ}$ for 10 ms pulses is shown in Fig. 2c. We can rule out a significant contribution to these changes from the spin torque due to the tunneling current passing through the MTJ, as this should add to or subtract from the spin Hall torque and hence should shift the switching curve as a whole along the current axis, which is not consistent with the behavior in Figs. 2b and 2c (see Section S2 of the Supplemental Information).

Because our measurements are conducted at room temperature, the magnitudes of the switching currents are reduced from their intrinsic values in the absence of thermal activation. To make quantitative comparisons to the predictions of Equations (1) and (2), we extrapolate to the fluctuation-free regime. We do this by measuring the switching currents as a function of pulse length (from 10 μs to 10 ms) while keeping the gate voltage pulse amplitude constant at $V_{MTJ} = \pm 400$ mV or 0 mV (Fig. 2d). The values plotted in Fig. 2d were determined from the average of at least ten scans for each pulse length. Within the framework of thermally-activated switching, the critical current is related to the pulse length $\tau$ through the relationship[26] $I_c = I_{c0}\left[1 - (k_B T / E)\ln(\tau / \tau_0)\right]$, where $I_{c0}$ is the zero-fluctuation critical current, $E$ is the energy barrier for thermal activation, and $\tau_0$ is the inverse of the attempt frequency assumed to be of order 1 ns. By fitting the data in Fig. 2d we determined values for $I_{c0}$ and the ratio of $E/k_B T$ corresponding to the different values of $V_{MTJ}$ (Table 1). We find that both $I_{c0}$ and $E/k_B T$ depend strongly on $V_{MTJ}$. Averaging the changes in $|I_{c0}|$ over the P-to-AP and AP-to-P transitions, we observe a total change $\Delta I_{c0} = 0.14 \pm 0.04$ mA between $V_{MTJ} = -400$ mV and 400 mV. Given the geometry of the sample and the resistivities $\rho_{Ta} = 190$ μΩ-cm and $\rho_{CoFeB} = 170$ μΩ-cm, this corresponds to a total change in zero-temperature current density $\Delta J_{c0} = (2.0 \pm 0.6) \times 10^6$ A/cm$^2$ in the Ta. Using Eq. (1) with the previously-determined[22] sample materials parameters $M_S =$ 1100 emu/cm$^3$, $\alpha$ = 0.021, and $\theta_{SH}$ = 0.15, we infer from these switching data a total shift in the effective demagnetization field $\Delta H_{demag}^{eff}$ = 570 ± 160 Oe, or $dH_{demag}^{eff}/dV_{MTJ}$ = 710 ± 200 Oe. This is in good quantitative agreement with the value we have previously determined for our multilayer[23], $d\left(H_{demag}^{eff}\right)/dV_{MTJ}$ = 730 ± 90 Oe/V, confirming the applicability of this aspect of Eqs. (1) and (2), as expected.

The dependence of $E/k_BT$ on $V_{MTJ}$ that we measure (Table 1) is somewhat more surprising. We find that $E/k_BT$ is reduced for both signs of $V_{MTJ}$, but much more strongly for negative bias. The fact that $E/k_BT$ decreases for both signs of $V_{MTJ}$ can explain why the measured room-temperature switching currents depend on $V_{MTJ}$ much more strongly for negative $V_{MTJ}$ than at positive values (Figs. 2b and 2c). For negative $V_{MTJ}$, the changes in $I_{c0}$ and $E/k_BT$ are reinforcing, in that both act to reduce the magnitudes of the switching currents, while for positive $V_{MTJ}$ the change in $I_{c0}$ with $V_{MTJ}$ acts to increase the switching current while the change in $E/k_BT$ favors a decrease, leading to a weaker overall effect. Part of the measured dependence of $E/k_BT$ on $V_{MTJ}$ may arise from an increase in $T$ caused by Ohmic heating, but the large asymmetry for $E/k_BT$ with respect to the sign of $V_{MTJ}$ indicates that the activation energy is voltage-dependent, too. This is confirmed by measurements of the dependence of the coercive magnetic field on $V_{MTJ}$ for field sweeps along the easy axis (see Section S3 of the Supplementary Information). We propose that the dependence of $E$ on $V_{MTJ}$ arises from a sample non-ideality; that because the overall effective demagnetization field in the thin $Co_{40}Fe_{40}B_{20}$ films is relatively small and depends on $V_{MTJ}$, the micromagnetic configuration of the free layer is spatially nonuniform and not completely in-plane, to an extent that depends on $V_{MTJ}$. This can affect the activation energy, e.g., by altering the energy cost for domain-wall-mediated reversal. Our data suggest that magnetization nonuniformities are particularly important in the P configuration, in that the MTJ resistance in this configuration depends on the applied magnetic field (Fig. 1b) and $E/k_BT$ is generally significantly less for P-to-AP switching than for the reverse. Furthermore, a spatially non-uniform magnetization state in the P configuration may explain the asymmetry between the values of $I_{c0}$ measured for the two switching directions (Table 1), because the efficiency of spin torque switching decreases as the average magnetization angle of the free layer shifts away from exactly parallel or antiparallel to the direction of the spin torque. In contrast to the results presented here, in our previous measurements with slightly thicker CoFeB free layers the values of $I_{c0}$ were symmetric for P-to-AP and AP-to-P switching[22].

For applications, the gate voltage should ideally be able to change the probability of switching fully between 0% and 100% for a given value of $I_{Ta}$. We provide an example of this capability measured to the 2% level in Figures 3a and 3b, using $V_{MTJ} = 0$ and -400 mV with 10 μs pulses. We see that for both P-to-AP and AP-to-P switching there is a window of current amplitude for which the switching probability is 100% for $V_{MTJ} = -400$ mV and 0% for $V_{MTJ} = 0$, so that $V_{MTJ}$ gates the switching process effectively. This is shown more directly in Fig. 3c; a voltage $V_{MTJ} = -400$ mV puts the device in the ON state for switching by the spin Hall torque from the Ta, while $V_{MTJ} = 0$ turns the switching OFF. Figure 3c equivalently demonstrates reliable implementation of the AND logic operation. Other logic operations can be achieved by varying background and pulse voltages or by cascading more than one spin Hall torque device[27].

While the demonstration shown in Fig. 3 corresponds to switching in the long-pulse, thermally-activated regime where the gating action benefits from the $V_{MTJ}$ dependence of both $I_{c0}$ and the activation energy $E$, with a properly optimized device structure the VMCA effect on $I_{c0}$ should by itself be strong enough to provide full gating of spin torque switching also in the short-pulse (100 ps – 10 ns) regime where thermal activation becomes unimportant (see Section S4 of the Supplementary Information). Note, as well, that although our demonstration utilizes MTJs with predominantly in-plane magnetized films, the same 3-terminal device strategy should be applicable to gating the spin-torque switching of perpendicularly magnetized devices. As has been shown previously[21], the spin Hall effect can be used as the switching source for magnetic moments that are out-of-plane, in which case the spin Hall critical currents are proportional to the perpendicular anisotropy field that can be controlled by $V_{MTJ}$ via the VCMA mechanism.

The most exciting consequence of our results is that they demonstrate the feasibility of improved circuit architectures for making magnetic memory and non-volatile logic technologies; for example, nonvolatile magnetic random access memory in the maximum-density cross-point geometry[28-30] shown schematically in Fig. 4. A major challenge for the successful implementation of cross-point memories using conventional spin torque switching with 2-terminal magnetic tunnel junctions is the issue of current flow via sneak paths[31] that leads to unintended switching events and increased power consumption during writing processes and decreased sensitivity during reading. In the circuit shown in Fig. 4, during the writing operation each memory cell can be addressed individually by applying a gate voltage to the MTJ from above while also applying a current through the Ta microstrip below to generate a spin Hall torque. MTJs with high impedance can be utilized ($R_{MTJ} \gg R_{Ta}$, where $R_{Ta}$ is the resistance of the Ta strips), effectively blocking all possible sneak paths for the writing current. For the reading operation, the effects of sneak currents can be ameliorated using a parallel reading scheme proposed in ref. 32. The overall benefit of the cross-point architecture depicted in Fig. 4 is that whereas spin-torque MRAM circuits currently in development require at least 1 transistor for every bit, the cross-point geometry can be made with only 1 transistor for every N bits in an array, thereby increasing the storage density significantly and reducing the complexity at the interface between the MTJs and CMOS circuit elements.

**Methods**

The samples were fabricated by first using magnetron sputtering to deposit on an oxidized Si substrate the multilayer substrate/Ta(6)/CoFeB(1.5)/MgO(1.2)/CoFeB(4)/Ta(5)/Ru(5) (thicknesses in nanometers). We used two steps of lithography and ion milling to define the MTJ and the bottom Ta microstrip depicted in Fig. 1a. The long axis of the MTJ was aligned perpendicular to the direction of current flow in the Ta strip to enable spin torque switching by the spin Hall effect with maximum efficiency. The thickness of the free CoFeB layer was chosen such that the perpendicular anisotropy from the CoFeB/MgO interface was sufficient to greatly reduced the effective demagnetization field[33], but was not strong enough to tilt the magnetic moment fully out of plane. No annealing was employed, in order to avoid further increase of the perpendicular magnetic anisotropy.

|         | AP-to-P       |         | P-to-AP       |         |
|---------|---------------|---------|---------------|---------|
|         | $I_{c0}$ (mA) | $E/k_BT$ | $|I_{c0}|$ (mA) | $E/k_BT$ |
| -400 mV | 0.43±0.03     | 18      | 0.87±0.03     | 18      |
| 0 mV    | 0.51±0.02     | 72      | 0.97±0.04     | 37      |
| 400 mV  | 0.55±0.02     | 53      | 1.03±0.03     | 31      |

**Table 1. Dependence on $V_{MTJ}$ for the zero-fluctuation current $I_{c0}$ and the energy barrier for thermal activation $E$, as determined from linear fits to the data in Fig 2d.**

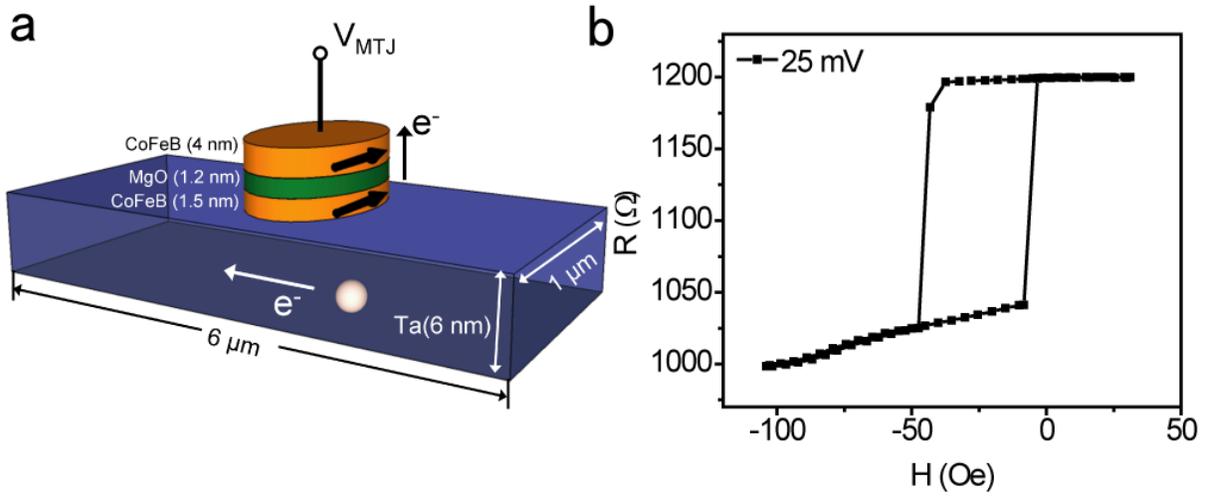

**Figure 1. Geometry and magnetic characterization of the 3-terminal MTJ device. a**, Schematic depiction of the device structure. The arrows label the direction of electron flow for positive biases. **b**, Magnetic minor loop of the MTJ measured at a small bias voltage ($V_{MTJ} = 25$ mV). A lead resistance from the Ta contact of 2 k$\Omega$ has been subtracted.

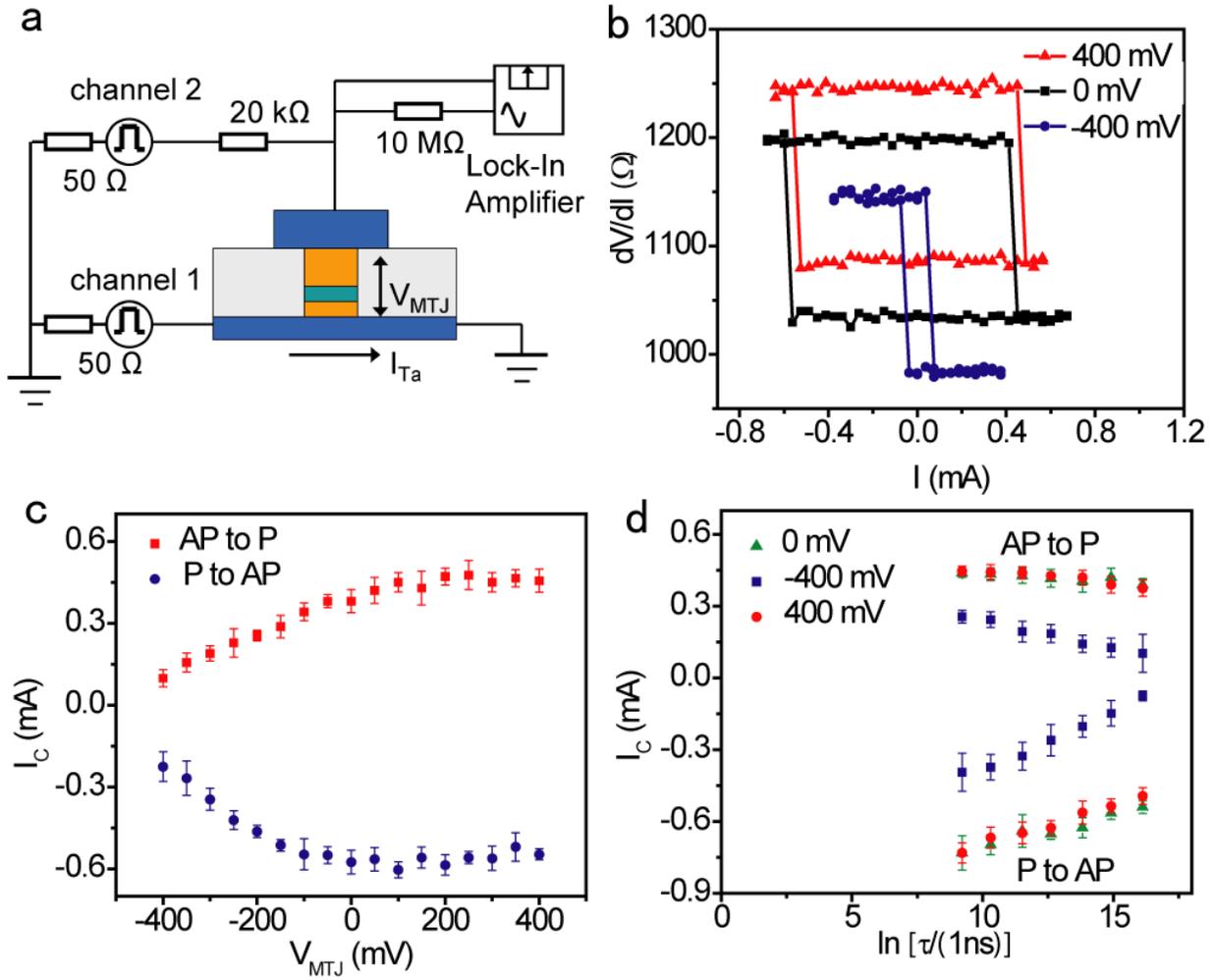

**Figure 2. Voltage gating of the spin Hall switching current. a,** Circuit used to apply simultaneous pulses to the MTJ and through the Ta microstrip. **b**, Spin Hall torque switching curves for different values of $V_{MTJ}$, for 10 ms pulses. The lead resistance from the Ta contact has been subtracted. The data for different $V_{MTJ}$ are offset vertically. c, The dependence of the room-temperature switching current $I_c$ on $V_{MTJ}$ for both P-to-AP (squares) and AP-to-P switching (circles). **d**, Pulse length dependence of the critical currents for spin Hall torque switching, for pulses from 10 μs to 10 ms.

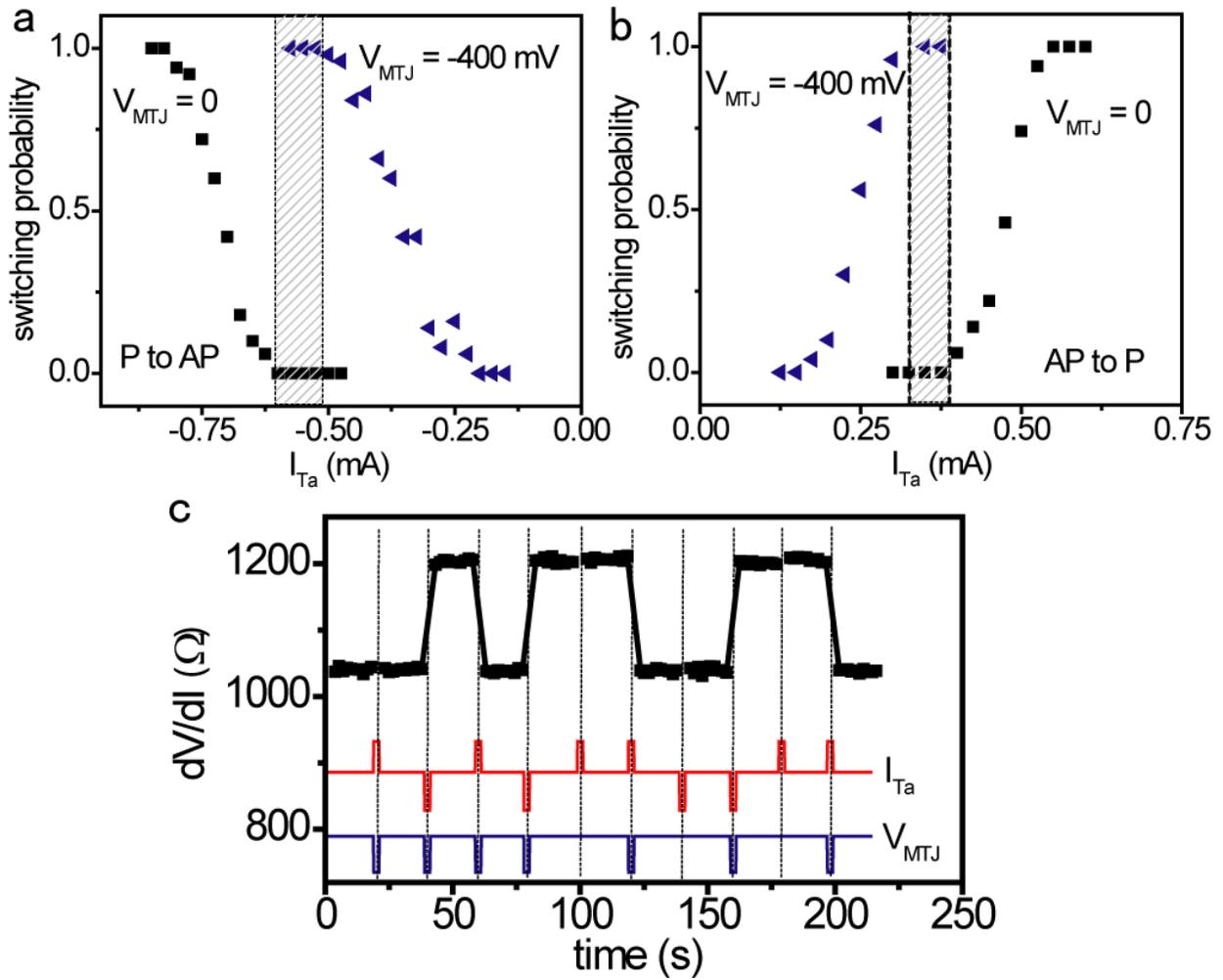

**Figure 3. Evaluating the effectiveness of gated spin Hall torque switching. a** and **b**, Comparison of the switching probability as a function of current amplitude in the Ta microstrip for the MTJ in the ON state ($V_{MTJ}$ = -400 mV) and the OFF state ($V_{MTJ}$ = 0 mV) for (**a**) P-to-AP switching and (**b**) AP-to-P switching. The pulse lengths are 10 µs and the switching probability for each value of $I_{Ta}$ was calculated from 50 attempts. The shaded areas represent regions where the gate voltage is effective in changing the switching probability from 0% to 100%. **c**, Gated spin Hall torque switching under a series of 10 µs pulses. $V_{MTJ}$ is switched between 0 mV and -400 mV, while $I_{Ta}$ is switched between 0 mA, 0.35 mA and -0.55 mA.

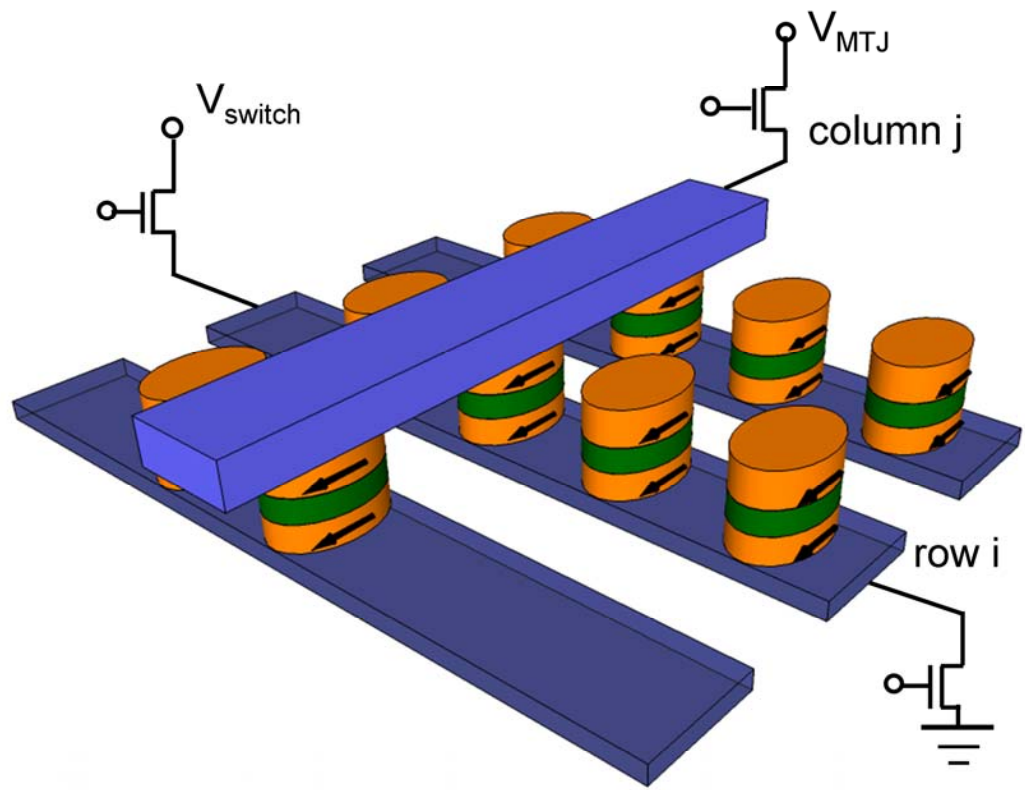

**Figure 4. Cross-point memory architecture enabled by gated spin Hall torque switching.**
(see Section S5 of the Supplementary Information for a more detailed discussion of the reading and writing operations).

# Supplementary Information for

# Gate voltage modulation of spin-Hall-torque-driven magnetic switching

Luqiao Liu[1], Chi-Feng Pai[1], D. C. Ralph[1,2] and R. A. Buhrman[1]

[1] Cornell University, Ithaca, NY 14853

[2] Kavli Institute at Cornell, Ithaca, NY 14853

## S1. Determination of the current and voltage across the MTJ

We use the equivalent circuit shown in Fig. S1 to determine the voltages and currents within our 3-terminal MTJ device. $R_{MTJ} \approx 1$ k$\Omega$ is the resistance of the MTJ (Fig. S2) and $R_{Ta} \approx 4$ k$\Omega$ is the resistance of the Ta microstrip. (The MTJ is located in the center of the Ta strip, so the left and right halves of the strip both have resistance $R_{Ta}/2$.) The MTJ is biased in series with a fixed resistance $R_S = 20$ k$\Omega$ by a voltage $V_2$, while a voltage $V_1$ is applied between the two ends of the Ta microstrip. $I_L$, $I_{MTJ}$ and $I_R$ represent the current flowing through the left part of the Ta strip, the MTJ, and the right part of the Ta strip. According to Kirchhoff's circuit laws, we have:

$$I_L + I_{MTJ} = I_R \tag{S1}$$

$$I_L R_{Ta}/2 + I_R R_{Ta}/2 = V_1 \tag{S2}$$

$$I_{MTJ}(R_S + R_{MTJ}) + I_R R_{Ta}/2 = V_2 \tag{S3}$$

The solution of Eqs. (S1)-(S3) yields $I_{MTJ} = (V_2 - 0.5V_1)/(R_S + R_{MTJ} + R_{Ta}/4)$, $I_L = V_1/R_{Ta} - 0.5 I_{MTJ}$, and $I_R = V_1/R_{Ta} + 0.5 I_{MTJ}$. The voltage across the MTJ is

$$V_{MTJ} = I_{MTJ} R_{MTJ} = R_{MTJ}(V_2 - 0.5V_1)/(R_S + R_{MTJ} + R_{Ta}/4). \tag{S4}$$

Because $I_{L,R} = V_1/(R_{Ta}) \mp 0.5 I_{MTJ}$, half of the current through the MTJ flows rightwards in the Ta strip and the other half flows leftwards. This indicates that for the region of the Ta located directly under the MTJ, the current $I_{MTJ}$ does not contribute to the spin Hall torque because the effects of right-going and left-going currents cancel. So, when calculating the spin torque from the spin Hall effect, we need to consider only the current generated by $V_1$, i.e., $I_{Ta} = V_1/R_{Ta}$.

When performing our experiments (*e.g.*, when taking data for Fig. 2 and Fig. 3 of the main text), we determine the value of $V_1$ needed to provide a given value $I_{Ta}$ using $V_1 = R_{Ta} I_{Ta}$ and then for this value of $I_{Ta}$ we calculate the value of $V_2$ needed to provide a given value of $V_{MTJ}$ using Equation (S4). In other words, we apply a correction so that the actual voltage dropped across the MTJ does not change as we vary $I_{Ta}$. This correction is not large; if it were left uncorrected, the largest current that we apply through the Ta strip, $I_{Ta}$ = 0.6 mA would lead to an extra tunneling current of ~50 μA or a tunneling current density of $1.4 \times 10^5$ A/cm$^2$, corresponding to a voltage drop of 50 mV across the MTJ.

**S2. Calculation of the spin currents from the tunneling current and the spin Hall effect**

The spin current associated with the tunneling current through the MTJ can be estimated using the zero-bias TMR ratio of 17%. Using the Julliere's model[S1] $\Delta R/R_P = 2 P_L P_R /(1 - P_L P_R)$ and assuming that the spin polarization for the left electrode $P_L$ and right electrode $P_R$ are equal, the spin polarization is ~ 0.3. This corresponds to a spin current density of $3 \times 10^5$ $\hbar/2e$ A/cm$^2$ for $|V_{MTJ}|$ = 400 mV. This is likely an overestimate because for large values of $|V_{MTJ}|$ the tunneling spin polarization may be decreased from the zero-bias value. This estimate of the spin current density due to the tunneling current can be compared to the spin current density due to the spin Hall effect for 0.5 mA in the Ta wire (near the switching current for $V_{MTJ}$ = 0):

$J_{S,Ta} = \theta_{SH} J_{e,Ta} \hbar / 2e \approx 1.2 \times 10^6 \ \hbar/2e$ A/cm$^2$.  Here we have used that the spin Hall angle in our Ta films is $\theta_{SH} \approx 0.15$ based on ref. [S2].  Therefore we expect the spin Hall torque to be significantly stronger than the spin torque due to the tunneling current.

As noted in the main text, the best evidence that the primary effect of the gate voltage on the switching currents is via voltage controlled magnetic anisotropy rather than spin torque from the tunneling current is that the spin torque from the tunneling current should simply add to the spin Hall torque so as to rigidly shift the hysteresis curve as a function of $I_{Ta}$ – shifting the switching currents for both P-to-AP and AP-to-P transitions in the same direction (so that the absolute values of the switching currents shift in opposite directions).  In contrast, the voltage controlled magnetic anisotropy should shift the absolute values of the switching currents in the same direction, as observed in the experiment.

## S3. Additional evidence for the effects of spatially non-uniform magnetic states

In the main text we argued that in addition to tuning the critical current for spin Hall switching by changing the strength of the perpendicular magnetic anisotropy, the applied gate voltage also alters the energy barrier for thermally-activated magnetic switching.  We suggested that this is because, by changing the perpendicular anisotropy, the gate voltage alters the degree of spatial non-uniformity in the free layer magnetization, so that some regions of the free layer tilt partially out of the sample plane.  Here we present additional evidence for the effects of this spatial nonuniformity.

Figure S3 shows the hysteresis loop for the MTJ resistance for an in-plane magnetic field swept parallel to the easy axis of the free layer, for a small MTJ voltage 15 mV.  If the magnetizations were spatially uniform, the resistance should be constant as a function of field in both the high (AP) and low (P) resistance states. However, we observe a strong field dependence in the P state for fields less than 100 Oe, indicating that the magnetization is not uniformly polarized in this regime.

The energy barrier for thermal activation should be related to the coercive magnetic field in the absence of thermal fluctuations, $H_{c0}$, approximately as $E \approx M_S H_{c0}/2$. Therefore if the gate voltage changes the energy barrier, as we claim, it should also affect the coercive field. Confirmation of this change is shown in Fig. S4. These data were taken at room temperature, so they are influenced by thermal fluctuations. Ohmic heating from $V_{MTJ}$ also likely reduces the magnitudes of the switching fields for $V_{MTJ}$ = -250 mV and 250 mV, relative to the low-bias (25 mV) curve.

## S4. Calculation for the modulation of the spin Hall torque switching currents by gate voltage in an optimized device, in the absence of thermal activation.

The experiments in the main text were conducted for pulse lengths from 10 μs to 10 ms, which are long enough that thermal activation contributes to magnetic reversal. This means that the gate voltage can act to modulate the spin Hall torque switching current via its effects on both the zero-fluctuation critical current $I_{c0}$ and the activation barrier $E$. However, for most applications, switching will be driven by pulses < 10 ns long, for which thermal activation has less effect. Therefore it is important to determine whether in this regime the gate voltage can effectively modulate the switching currents just through its influence on $I_{c0}$. We assume that the perpendicular magnetic anisotropy of the free layer is optimized so that $H_{demag}^{eff} \approx 1$ kOe (ref. [S3]) with $d(H_{demag}^{eff})/dV_{MTJ}$ = 710 Oe/V, the value we determined for our sample. We also assume that $H_c$ is large enough to maintain the thermal stability of the free layer, but $H_c \ll H_{demag}^{eff}$. Using the parameter values[S2] $M_S$ = 1100 emu/cm$^3$, $t_{free}$ = 1.5 nm, $\alpha$ = 0.021, and $\theta_{SH}$ = 0.15, Eq. (2) of the main text yields $|J_{c0}|$ = 9.6 × 10$^6$ A/cm$^2$ for $V_{MTJ}$ = 500 mV and $|J_{c0}|$ = 4.5 × 10$^6$ A/cm$^2$ for $V_{MTJ}$ = -500 mV. This factor of two variation is larger than the typical width of the distribution for the switching current density in spin torque devices (see, e.g., ref. [S4]), so that the effect of the voltage-controlled anisotropy on $J_{c0}$ should indeed be sufficient by itself to achieve full modulation of spin Hall torque switching in optimized devices.

**S5. Bias configurations for the writing and reading operations in the gated spin Hall torque cross-point memory**

For the writing operation, the transistor at the chosen column and the pair of the transistors at the two ends of the chosen row are set to be ON while all of the other transistors are set to be OFF (Fig. S5a). $V_{switch}$ is chosen to be positive or negative depending on which final state is desired for the MTJ. Information is then written into the MTJ represented by the red cross. For reading, one could use the parallel reading scheme of ref. [S5] (Fig. S5b). The transistors for all of the columns and the transistor at the right end of the chosen row are set to be ON. All of the other transistors are set to be OFF. Therefore, all of the column lines are set at the same voltage level +V. Information is read in a parallel way from all the MTJs on the same row by measuring the currents flowing in the column lines.

**Supplemental references**

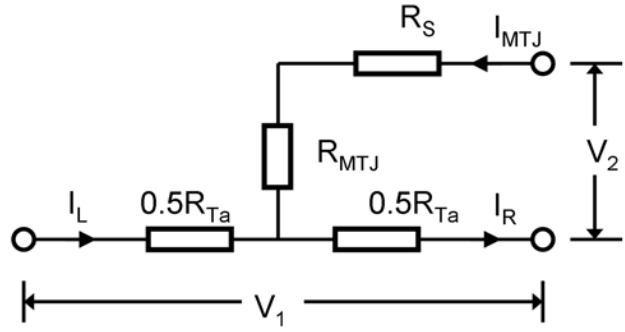

**Figure S1. Equivalent circuit for the device studied in Fig 2a of the maintext.** $R_{Ta}$ is the Ta strip resistance. $R_{MTJ}$ is the resistance of the MTJ. $R_S = 20$ k$\Omega$ is a fixed resistance connected in series with the MTJ. $V_1$ and $V_2$ are the applied voltages.

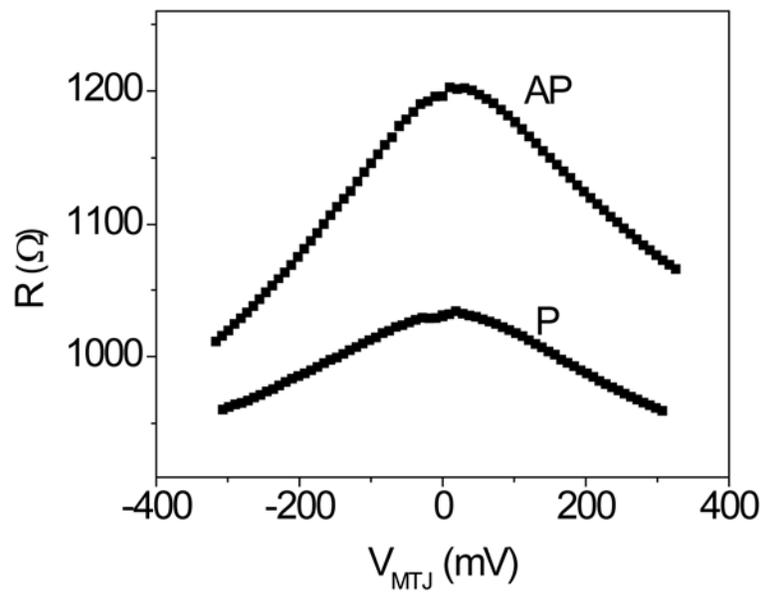

Figure S2. Dependence of the MTJ resistance on the MTJ voltage for the antiparallel and (approximately) parallel configurations.

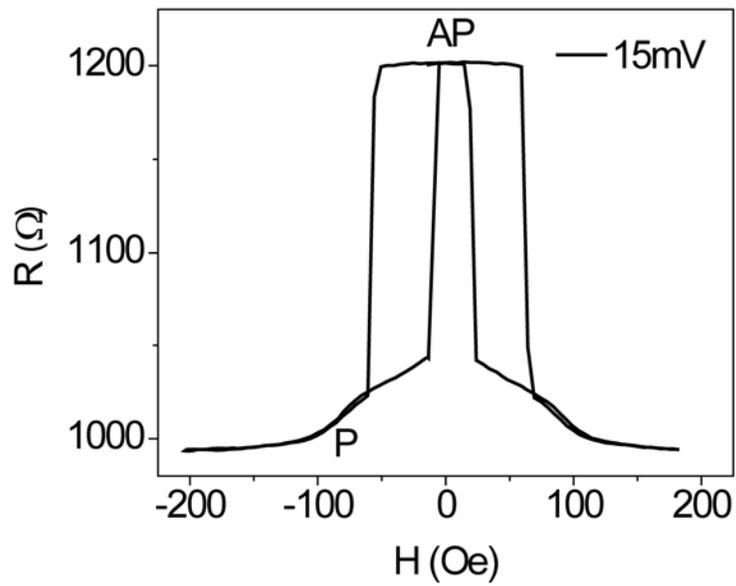

**Figure S3. Major loop of the device measured from the resistance across the MTJ.** The slope associated with the P state indicates that the equilibrium position for the free layer magnetization is not completely collinear with the fixed layer when external field of $H = -25$ Oe is applied.

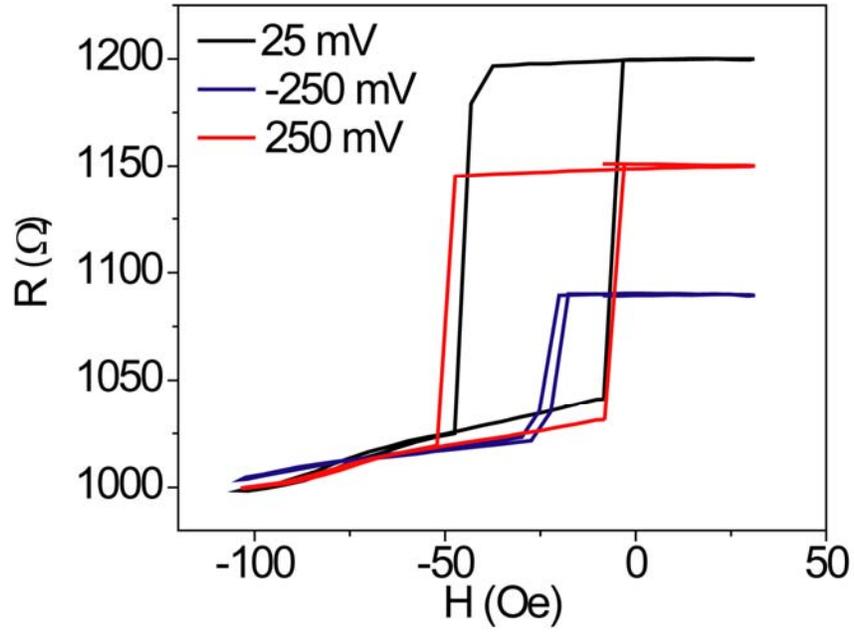

**Figure S4. Dependence of the coercive field on gate voltage.** Magnetic minor loop of the MTJ under bias voltages of $V_{MTJ} = 25$ mV and $\pm 250$ mV. For ease of comparison, switching curves for $V_{MTJ} = \pm 250$ mV are shifted vertically to make their resistances in the parallel state the same as for $V_{MTJ} = 25$ mV.

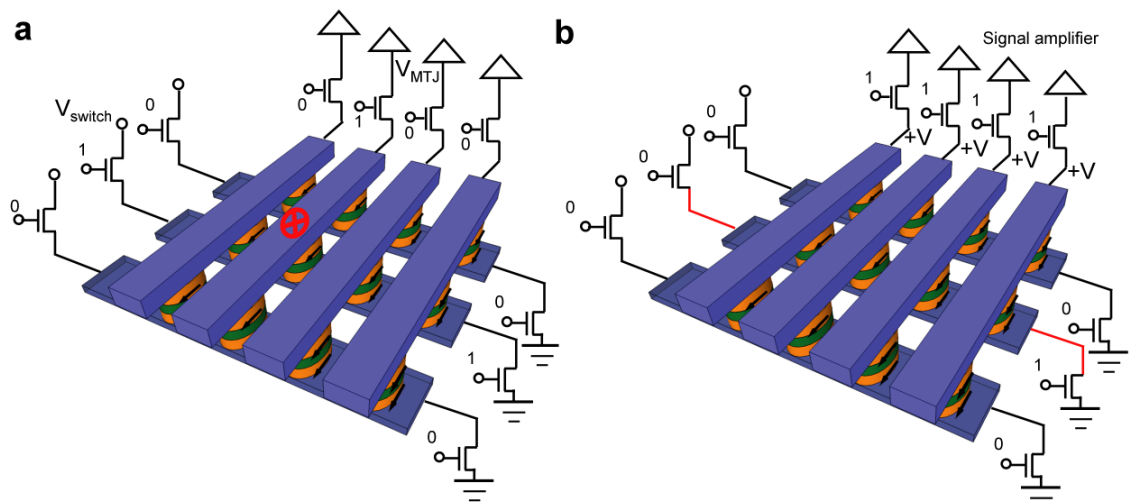

**Figure S5. Schematic illustration of the reading and writing operations for the cross-point memory.**